\documentclass[prl,twocolumn,showpacs,preprintnumbers,amsmath,amssymb]{revtex4-1}
\usepackage{graphicx}
\usepackage{dcolumn}
\usepackage{bm}

\newcommand{\dd}{\mathrm{d}}

\begin{document}

\title{Effect of bond length fluctuations on crystal nucleation of hard bead chains}

\author{Ran Ni}
\email{r.ni@uu.nl}
\author{Marjolein Dijkstra}
\email{m.dijkstra1@uu.nl}
\affiliation{%
Soft Condensed Matter, Debye Institute for NanoMaterials Science, Utrecht
University, Princetonplein 5 3584CC,
Utrecht, The Netherlands
}%

\date{\today}

\begin{abstract}
We study the effect of bond length fluctuations on the nucleation rate and crystal morphology of linear and cyclic chains  of flexibly connected hard spheres using  extensive molecular dynamics simulations.  For bond length fluctuations as small as a tenth of the bead diameter, the relaxation and crystallization resemble that of disconnected spheres.  For shorter bond lengths and  chains with $<10$ beads the nucleation rates depend sensitively on bond length, number of beads per chain, and chain topology, while for longer chains the nucleation rates are rather independent of chain length. Surprisingly, we find a nice data collapse of the nucleation rate as a function of number of bonds per sphere in the system, independent of bead chain composition, chain length and topology. Hence, the crystal nucleation rate of bead chains can be enhanced by adding monomers to the system. We also find that the resulting  crystal morphologies of bead chains {\em with bond fluctuations} resemble closely those for hard spheres including structures with five-fold symmetry.
\end{abstract}

\maketitle

Although crystal nucleation from a supersaturated fluid is one of the most fundamental processes during solidification, the mechanism is still far from being well understood. Even in a relatively simple model  system of pure hard spheres, the nucleation rates obtained from Monte Carlo (MC) simulations using the umbrella sampling technique differ by more than 6 orders of magnitude from those measured in experiments~\cite{auer2001}. This discrepancy in the nucleation rates has led to intense ongoing debates in the past decade on the reliability of various techniques as employed in simulations and experiments  to obtain the nucleation rates~\cite{tanaka2010,filion2010}. Recently, it was shown that the theoretical prediction of the nucleation rates  for hard spheres is consistent for three widely used rare-event techniques, despite the fact that the methods treat the dynamics very differently~\cite{filion2010}.
Moreover, the structure of the resulting crystal nuclei as obtained from the different simulation techniques all agreed and showed that the nuclei consist of approximately 80$\%$ face-centered-cubic-like particles. The predominance of face-centered-cubic-like particles in the critical  nuclei is unexpected, as the free energy difference between the bulk face-centered-cubic (fcc) and hexagonal-close-packed (hcp) phases is about 0.001 $k_B T$ per particle, and one would thus expect to find a random-hexagonal-close-packed (rhcp) crystal phase  \cite{bolhuis1997}.  More surprisingly, simulation studies showed that the subsequent growth of these critical nuclei resulted in a range of crystal morphologies with a predominance of multiple twinned structures exhibiting in some cases structures with a five-fold symmetry  \cite{malley2003,karayiannis2011}.  Such structures are intriguing as the fivefold symmetry is incompatible with space-filling periodic crystals. Moreover, the formation mechanism of these fivefold structures is still unknown. Bagley speculated that the fivefold structures are due to the growth of fivefold local structures (a decahedral or pentagonal dipyramid cluster of spheres) \cite{anikeenko2007} that are already present in the supersaturated fluid phase \cite{bagley1970}. Another mechanism that has been proposed is that these multiple twinned structures with a fivefold symmetry originates from nucleated fcc domains that are bound together by stacking faults \cite{anikeenko2007}. For instance, five tetrahedral fcc domains can form a cyclic multiple twinned structure with a pentagonal pyramid shape. A recent event-driven Molecular Dynamics simulation study on hard spheres showed, however, no correlation between the fivefold local clusters that are already present  in the supersaturated fluid and the multiple twinned structures in the final crystal phase~\cite{karayiannis2011}. Hence, it was concluded that crystalline phases with multiple stacking directions may possess fivefold structures, whereas crystals with a unique stacking direction do not show any five-fold symmetry patterns. These authors also succeeded to study the crystallization of tangent hard-sphere chains  by introducing several  clever (unphysical) chain-connectivity altering MC moves~\cite{mcmethod}.  Surprisingly, they observed that tangent hard-sphere chains (no bond fluctuations) never formed crystalline structures  with a fivefold symmetry as  the chain connectivity prohibits the formation of twinned structures and forces the crystals to grow in a single stacking direction ~\cite{karayiannis2011,karayiannis2009}.

In this Letter, we study the effect of bond length fluctuations on the nucleation rate and crystal morphology  of flexibly connected hard spheres.  These hard-sphere chains with bond fluctuations can serve as a simple model for granular ball chains~\cite{nagel2009}, colloidal bead chains~\cite{raopolymer} or (short) polymeric systems.  A better understanding of the behavior of these bead chains may shed light on the glass transition and crystallization of polymers. In fact, it has been shown recently that random packings of granular ball chains show striking similarities with the  glass transition in polymers~\cite{nagel2009}. Moreover, the scaling dependence of a number of physical characteristics
on packing densities at concentrations which still remain unreachable in simulations of atomistic or
coarse-grained polymer models has been studied by employing this polymeric hard-sphere chain model~\cite{scale1,scale2}.
We also mention that recently, a colloidal model system of bead chains has been realized consisting of colloidal spheres  that are bound together with  ``flexible linkers'' ~\cite{raopolymer}.

We consider a system of $M$ bead chains consisting of $N$ identical hard spheres with diameter $\sigma$ in a volume $V$. The hard-sphere beads are connected by flexible bonds with a bond energy  $U_{bond}(r_{ij})$ given by
\begin{equation}
 \frac{U_{bond}(r_{ij})}{k_BT} = \left\{
\begin{array}{cc}
 0 & \sigma <r_{ij} < \sigma + \delta \\
 \infty	& \mathrm{otherwise}
\end{array}\right.
\end{equation}
where $r_{ij}$ is the center-to-center distance between two connected spherical beads $i$ and $j$, $\delta$ is the maximum bond length, $k_B$ the Boltzmann constant, and $T$ the temperature. The maximum bond length $\delta$  varies from $0$ to $0.1\sigma$ in our simulations, such that $\delta = 0$ corresponds to a freely jointed chain of tangent hard spheres.  We  employ event driven molecular dynamics (EDMD) simulations~\cite{rapaportpaper,hartmannjcp} instead of the extensive MC method~\cite{mcmethod} to mimic the dynamics of  granular and colloidal bead chains with flexible bonds.  Since the pair potentials between all spheres are discontinuous, the pair interactions only change when the beads collide or when the maximum bond length is reached. Hence,  the particles move in straight lines (ballistically) until they encounter another particle or reach the maximum bond length. The particles then perform an elastic collision. These collisions are identified and handled in order of occurrence using an EDMD simulation~\cite{rapaportmd}.

We  study the effect of bond length fluctuations on the nucleation of bead chains at a packing fraction $\eta=0.55$, which is at the upper boundary of  the nucleation  regime of hard spheres. At $\eta=0.55$, the nucleation time exceeds the relaxation time, which enables us to equilibrate the metastable fluid and to measure the nucleation rate. In order to obtain the initial configuration for our EDMD simulations, we use the  Lubachevsky-Stillinger algorithm~\cite{LSA} to grow the beads of the chains in the simulation box to the packing fraction of interest with a very fast speed, i.e.,  $\dd\sigma(t)/\dd t=0.01\sigma/\tau$ where $\sigma(t)$ is the size of spheres at time $t$ with $\sigma$ the target sphere size and $\tau=\sigma\sqrt{m/k_BT}$ the MD time scale.
We monitor the relaxation of the initial configurations by comparing the bond angle distributions with those of previous work \cite{karayiannis2010} and by calculating the long-time self diffusion coefficient $D_L$ of the beads. We find that $D_L$  increases linearly with   $1/N$ as predicted by the Rouse model~\cite{bookdoi}, and is rather independent of the bond length $\delta$, at least for the range of values that we studied~\cite{supinf}. 
In order to exclude the effect of dynamics on the nucleation rates~\cite{filion2010,ran2010jcp}, we  use the long-time diffusion time of the beads, $\tau_L=1/6D_L$, as the unit of time in the nucleation rate.
Starting from the metastable fluid phase, an EDMD simulation is employed to evolve the system until a spontaneous nucleation event occurs.
The nucleation rate is then given by $I = 1/\langle t \rangle V$, where $\langle t \rangle$ is the average waiting time before a nucleation event occurs in a system of volume $V$. In order to identify the crystalline clusters in the fluid phase, we employ the local bond-order parameter analysis~\cite{bop,ran2010jcp}.

\begin{figure}
 \includegraphics[width=0.48\textwidth]{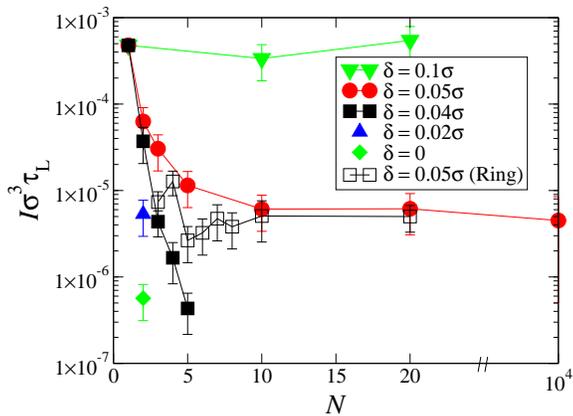}
  \caption{\label{fig2}(Color online) Nucleation rates $I \sigma^3 \tau_L$ of linear hard-sphere polymers with maximum bond length $\delta=0.1 \sigma$ (green triangles),  $0.05 \sigma$ (solid circles),  $0.04 \sigma$  (solid squares), and of ring-like hard-sphere polymers with maximum bond length $\delta=0.05 \sigma$ and chain length (open squares). For comparison, we also plot the nucleation rate for hard dumbbells with a maximum bond length $\delta = 0.02 \sigma$ (blue triangles) and $\delta=0$ (diamonds). }
\end{figure}

\begin{figure}
 \includegraphics[width=0.4\textwidth]{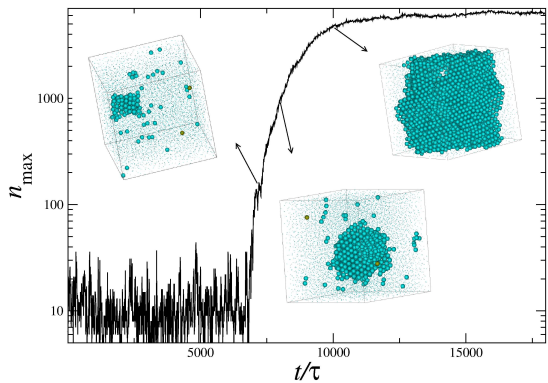}
  \caption{\label{fig1n}(Color online) Size of the largest crystalline cluster $n_{\max}$ as a function of simulation time for a single linear hard-sphere polymer with chain length $N=10^4$. The insets are snapshots at $t=7060\tau$, $8000\tau$ and $10000\tau$, respectively, where only the solid-like beads are shown, and the two dark yellow spheres are the two ends of the chain.}
\end{figure}

We calculate the crystal nucleation rates in systems consisting of $\simeq 10^4$ beads  for various linear and ring-like hard-sphere polymers for chain length $1 \leq N \leq 20$ and various bond length $\delta$. We mention here that we did not exclude inter-ring catenations in the configurations of ring polymers, which may occur for rings with $N \geq 5$. Figure~\ref{fig2} shows that for $\delta$ as small as $0.1 \sigma$, the nucleation rate resembles that of loose hard spheres. For $\delta < 0.1 \sigma$, we find that the nucleation rate   decreases by several orders of magnitude with decreasing $\delta$ and  increasing chain length $N$, but remains constant for sufficiently long  polymers, i.e., $N \ge 10$. To check this surprising result, we also determined the nucleation rate of a single polymer of length $N=10^4$ and $\delta = 0.05 \sigma$, where all the beads are doubly connected except the two end beads.  We observe that the system remains in the fluid phase for $\simeq 7000\tau$, before a critical nucleus of $\sim 100$ beads forms in the middle of the chain, which subsequently grows further until the whole system is crystalline. In Fig.~\ref{fig1n}, we show the size of the largest crystalline cluster as a function of simulation time from a typical MD trajectory. Additionally, we find that the nucleation rate does not decrease significantly for $N=10^4$, which is highly unexpected as the beads can only move collectively.

Additionally, we did not observe spontaneous nucleation within our  simulation times   for linear bead chains with   $\delta=0.04 \sigma$ and  $N>5$, which is to be expected as the  bead chains  become too frustrated on an fcc lattice at $\eta=0.55$. At this density, the lattice spacing is much larger  than the sum of the maximum bond length and bead diameter and thus the nucleation is strongly inhibited by geometrical frustration.  However, for higher packing fraction (lattice spacing similar to $\delta+\sigma$), the nucleation barrier vanishes at such a high supersaturation and crystallization sets in immediately: a nucleation rate is here undefined. For lower $\eta$, the nucleation times for bead chains exceed rapidly our simulation times, and again the rate cannot be determined.  Finally, we also determine the nucleation rates for cyclic bead-chains (ring polymers) with bond length $\delta =0.05\sigma$. Figure~\ref{fig2} shows  that the nucleation rate of ring polymers does not decrease monotonically with $N$ and is always lower than for linear polymers with the same length. However, the  difference in nucleation rate  is  small for $N \geq 10$.

\begin{figure}
 \includegraphics[width=0.5\textwidth]{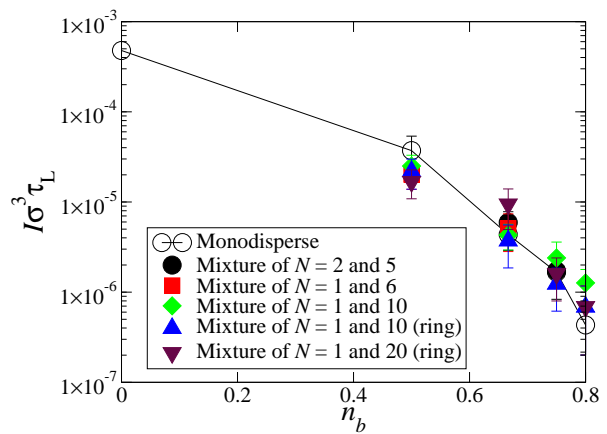}
  \caption{\label{fig3}(Color online) Nucleation rates  $I \sigma^3 \tau_L$ of linear hard-sphere polymers with chain length $N=1, 2, 3,$ and 5 and of binary mixtures of linear hard-sphere polymers with chain length $N=2$ and 5, $N=1$ and 6, and $N=1$ and 10, as a function of the average number of bonds per bead $n_b$. The composition of the mixture was chosen such that the value for $n_b$ matches with one of the values for the pure systems. The maximum bond length equals $\delta=0.04\sigma$ for all bead chains. }
\end{figure}

As mentioned above, our results show that the nucleation rates of linear polymers decreases with chain length $N$. One may argue that the nucleation rate is largely determined by the chain connectivity or the average number of bonds per sphere in the system.  In Fig. 3, we plot the nucleation rate for linear bead chains with maximum bond length $\delta =0.04\sigma$ and chain length $N=1, 2, 3,4,$ and 5, which correspond to an average number of bonds per bead of $n_b=(N-1)/N=0, 1/2, 2/3, 3/4$ and $4/5$. In order to investigate the effect of average number of bonds per bead in the system on the nucleation rate, we also perform simulations for binary mixtures of linear and ring polymers with different chain lengths and bond length $\delta =0.04\sigma$. We consider mixtures of linear chains with length $N=2$ and 5, $N=1$ and 6, and $N=1$ and 10, and a mixture of monomers and ring polymers with $N=10$ and 20. The composition of the mixture is chosen such that the value of $n_b$ matches with one of the values for the pure systems. We compare the nucleation rates for the pure and binary systems in  Fig.~\ref{fig3}, and observe a nice data collapse, suggesting that  the nucleation rate is determined by the number of bonds per sphere in the system, and the frustration imposed by the chain connectivity in these systems. We wish to remark here that no nucleation was observed in pure systems of linear or ring polymers with $\delta = 0.04\sigma$ and $N \ge 10$  within our simulation times. Thus the addition of monomers enhances significantly  the nucleation of polymers.

\begin{figure}
 \includegraphics[width=0.5\textwidth]{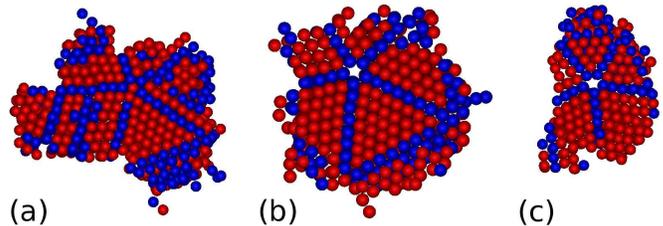}
  \caption{\label{fig4}(Color online) Typical configurations of the crystal structures for linear hard-sphere chains  with chain length $N=20$ (a) and for ring-like polymers with $N=3$(b) and $N=5$(c). Only crystalline spheres are shown here.  The blue and red spheres are hcp-like and fcc-like particles, respectively.}
\end{figure}

Additionally, we  investigate the structure of the resulting critical nuclei and the full-grown crystals by calculating the averaged local bond order parameters $\overline{q}_4$ and $\overline{w}_4$ for each sphere $i$ that has $N_b(i) \geq 10$ neighbours. This analysis allows us to check whether a bead is fcc-like or hcp-like~\cite{ran2010jcp,lechner}. We find that the critical nuclei contain more fcc-like than hcp-like particles, which is very similar to the critical nuclei observed in hard-sphere~\cite{filion2010} and hard-dumbbell nucleation~\cite{ran2010jcp}. In addition, we find that the full-grown crystals  display a range of crystal morphologies including random-stacked hexagonal close-packed crystals, and structures with a five-fold symmetry pattern for all polymer systems that we considered, even for ring-like polymers with chain length as small as  $N=3$, which were not observed in the crystal of freely jointed chains of tangent hard spheres \cite{karayiannis2011}.  Exemplarily, Fig.~\ref{fig4} shows typical configurations of these five-fold symmetry patterns formed by linear hard-sphere chains of length $N=20$ and  $\delta =0.05\sigma$, and cyclic bead chains of  $N=3$ and $5$ and $\delta =0.05\sigma$. We note that the crystal structures resemble closely those observed in MD simulations of  hard spheres~\cite{malley2003,karayiannis2011}.  As the crystal morphology is mainly determined by the crystallization kinetics rather than the bulk and surface contributions to the free energy of the nucleus, it is tempting to speculate  that the crystallization dynamics of hard bead chains {\em with bond length fluctuations} is similar to that of hard spheres, which is different from tangent hard-sphere chains (no bond length fluctuations), which crystallize into stacking faulted layered crystals with unique stacking directions and no five-fold symmetry patterns~\cite{karayiannis2009,karayiannis2010}.

\begin{figure}
 \includegraphics[width=0.4\textwidth]{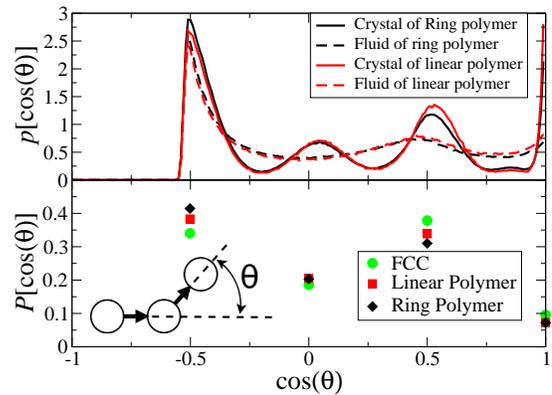}
   \caption{\label{fig5}(Color online) Bond angle distribution function  $p(\cos \theta)$ for a supersaturated fluid and crystal  of linear and cyclic hard-sphere polymers with  chain length $N=10$ and $\delta = 0.05\sigma$(top). Integration of the peaks of $p(\cos \theta)$  between the two neighboring local minima (bottom). For comparison, we also plot the bond angle distribution for self-avoiding random walks consisting of 10 steps on an fcc crystal lattice (bottom). The inset shows the definition of the angle $\theta$.}
\end{figure}

Furthermore, we determine the bond angle distribution function $p( \cos \theta)$ in order to quantify the  distribution of bond angles  between two neighboring polymer bonds in the supersaturated fluid and crystal nuclei. Figure~\ref{fig5} shows $p( \cos \theta)$  for systems consisting of linear and cyclic bead chains with $N=10$. For the fluid phase, $p( \cos \theta)$  displays two peaks at $\cos \theta = -0.5$ and $0.5$, i.e., $\theta = 120$ and  60 degrees, which corresponds with the most frequent three  particle structures observed in random packings of spheres~\cite{karayiannis2010}. However, $p( \cos \theta)$ of the crystal structures exhibits  four pronounced peaks located around $\cos \theta = -0.5, 0, 0.5, 1.0$ and a smaller peak at  $\cos \theta = \sqrt{3}/2$ , which corresponds with $\theta = 120, 90, 60, 0$ and 30 degrees, respectively~\cite{karayiannis2010}. In order to compare $p( \cos \theta)$  with that for a self-avoiding random walk on an fcc crystal lattice, we integrate the four pronounced peaks of $p(\cos \theta)$  between the two neighboring local minima.  The results are shown in Fig.~\ref{fig5} together with the bond angle distribution function for a self-avoiding random walk on an fcc crystal lattice. We find that $p( \cos \theta)$  for  crystals of  linear and cyclic polymers agree well with that of a self-avoiding random walk \cite{karayiannis2009}. Hence, $p( \cos \theta)$  seems not to be affected by the chain connectivity of the polymer chains. Free energy calculations show indeed that the stable solid of freely jointed hard-sphere chains is an aperiodic crystal phase, where the spheres are positioned on an fcc lattice with the bonds randomly oriented \cite{malanoski}.

In conclusion, we studied the effect of bond length fluctuations on the nucleation rate and crystal morphology of linear and cyclic hard-sphere chains  by varying the maximum bond length parameter $\delta$. For sufficiently large $\delta$, i.e. $\sim 0.1 \sigma$, the relaxation and crystallization resemble that of disconnected spheres. Hence, we find nucleation rates and crystal morphologies similar to that observed for hard spheres.  For  bond length values $0 <\delta <0.1 \sigma$,  our results show on the one hand that the nucleation rates decreases significantly with $\delta$, but on the other hand, the crystal morphology of bead chains with floppy bonds resemble closely that of loose hard spheres  \cite{malley2003,karayiannis2011}.  Apparently,  the bond constraints slows down the relaxation  dynamics as the crystallization is frustrated by the chain connectivity, but does not alter the crystallization kinetics. We thus find that  the resulting crystal nuclei of bead chains with bond  fluctuations possess a range of crystal morphologies including structures with a five-fold symmetry similar to hard-sphere crystals, while tangent hard-sphere chains  (no bond length fluctuations) crystallize into stacking faulted layered crystals with unique stacking directions and no five-fold symmetry patterns~\cite{karayiannis2009,karayiannis2010}. Hence, the crystallization kinetic pathways of bead chains depends strongly on bond length fluctuations. Additionally,  we find a nice data collapse of the nucleation rate as a function of number of bonds per sphere in the system, independent of bead chain composition, chain length and topology. The crystal nucleation rate can thus be enhanced  by the addition of monomers.
\begin{acknowledgments}
Financial support of a VICI grant from the Nederlandse Organisatie voor Wetenschappelijk Onderzoek (NWO) is acknowledged.
\end{acknowledgments}


\begin{thebibliography}{10}%
\makeatletter
\providecommand \@ifxundefined [1]{%
 \ifx #1\undefined \expandafter \@firstoftwo
 \else \expandafter \@secondoftwo
\fi
}%
\providecommand \@ifnum [1]{%
 \ifnum #1\expandafter \@firstoftwo
 \else \expandafter \@secondoftwo
\fi
}%
\providecommand \enquote [1]{``#1''}%
\providecommand \bibnamefont  [1]{#1}%
\providecommand \bibfnamefont [1]{#1}%
\providecommand \citenamefont [1]{#1}%
\providecommand\href[0]{\@sanitize\@href}%
\providecommand\@href[1]{\endgroup\@@startlink{#1}\endgroup\@@href}%
\providecommand\@@href[1]{#1\@@endlink}%
\providecommand \@sanitize [0]{\begingroup\catcode`\&12\catcode`\#12\relax}%
\@ifxundefined \pdfoutput {\@firstoftwo}{%
 \@ifnum{\z@=\pdfoutput}{\@firstoftwo}{\@secondoftwo}%
}{%
 \providecommand\@@startlink[1]{\leavevmode}%
 \providecommand\@@endlink[0]{}%
}{%
 \providecommand\@@startlink[1]{%
  \leavevmode
  \pdfstartlink
   attr{/Border[0 0 1 ]/H/I/C[0 1 1]}%
   user{/Subtype/Link/A<</Type/Action/S/URI/URI(#1)>>}%
  \relax
 }%
 \providecommand\@@endlink[0]{\pdfendlink}%
}%
\providecommand \url  [0]{\begingroup\@sanitize \@url }%
\providecommand \@url [1]{\endgroup\@href {#1}{\urlprefix}}%
\providecommand \urlprefix [0]{URL }%
\providecommand \Eprint[0]{\href }%
\@ifxundefined \urlstyle {%
  \providecommand \doi [1]{doi:\discretionary{}{}{}#1}%
}{%
  \providecommand \doi [0]{doi:\discretionary{}{}{}\begingroup
  \urlstyle{rm}\Url }%
}%
\providecommand \doibase [0]{http://dx.doi.org/}%
\providecommand \Doi[1]{\href{\doibase#1}}%
\providecommand \bibAnnote [3]{%
  \BibitemShut{#1}%
  \begin{quotation}\noindent
    \textsc{Key:}\ #2\\\textsc{Annotation:}\ #3%
  \end{quotation}%
}%
\providecommand \bibAnnoteFile [2]{%
  \IfFileExists{#2}{\bibAnnote {#1} {#2} {\input{#2}}}{}%
}%
\providecommand \typeout [0]{\immediate \write \m@ne }%
\providecommand \selectlanguage [0]{\@gobble}%
\providecommand \bibinfo [0]{\@secondoftwo}%
\providecommand \bibfield [0]{\@secondoftwo}%
\providecommand \translation [1]{[#1]}%
\providecommand \BibitemOpen[0]{}%
\providecommand \bibitemStop [0]{}%
\providecommand \bibitemNoStop [0]{.\EOS\space}%
\providecommand \EOS [0]{\spacefactor3000\relax}%
\providecommand \BibitemShut [1]{\csname bibitem#1\endcsname}%
\bibitem{auer2001}%
  \BibitemOpen
  \bibfield{author}{%
  \bibinfo {author} {\bibfnamefont{S.}~\bibnamefont{Auer}}\ and\ \bibinfo
  {author} {\bibfnamefont{D.}~\bibnamefont{Frenkel}},\ }%
  \bibfield{journal}{%
  \bibinfo {journal} {Nature (London)}\ }%
  \textbf{\bibinfo {volume} {409}},\ \bibinfo {pages} {1020} (\bibinfo {year}
  {2001})%
  \bibAnnoteFile{NoStop}{auer2001}%
\bibitem{tanaka2010}%
  \BibitemOpen
  \bibfield{author}{%
  \bibinfo {author} {\bibfnamefont{T.}~\bibnamefont{Kawasaki}}\ and\ \bibinfo
  {author} {\bibfnamefont{H.}~\bibnamefont{Tanaka}},\ }%
  \bibfield{journal}{%
  \bibinfo {journal} {Proc. Nat. Acad. Sci. (USA)}\ }%
  \textbf{\bibinfo {volume} {107}},\ \bibinfo {pages} {14036} (\bibinfo {year}
  {2010})%
  \bibAnnoteFile{NoStop}{tanaka2010}%
\bibitem{filion2010}%
  \BibitemOpen
  \bibfield{author}{%
  \bibinfo {author} {\bibfnamefont{L.}~\bibnamefont{Filion}}, \bibinfo {author}
  {\bibfnamefont{M.}~\bibnamefont{Hermes}}, \bibinfo {author}
  {\bibfnamefont{R.}~\bibnamefont{Ni}},\ and\ \bibinfo {author}
  {\bibfnamefont{M.}~\bibnamefont{Dijkstra}},\ }%
  \bibfield{journal}{%
  \bibinfo {journal} {J. Chem. Phys.}\ }%
  \textbf{\bibinfo {volume} {133}},\ \bibinfo {pages} {244115} (\bibinfo {year}
  {2010})%
  \bibAnnoteFile{NoStop}{filion2010}%
\bibitem{bolhuis1997}%
  \BibitemOpen
  \bibfield{author}{%
  \bibinfo {author} {\bibfnamefont{P.}~\bibnamefont{Bolhuis}}, \bibinfo
  {author} {\bibfnamefont{D.}~\bibnamefont{Frenkel}}, \bibinfo {author}
  {\bibfnamefont{S.}~\bibnamefont{Mau}},\ and\ \bibinfo {author}
  {\bibfnamefont{D.}~\bibnamefont{Huse}},\ }%
  \bibfield{journal}{%
  \bibinfo {journal} {Nature}\ }%
  \textbf{\bibinfo {volume} {388}},\ \bibinfo {pages} {235} (\bibinfo {year}
  {1997})%
  \bibAnnoteFile{NoStop}{bolhuis1997}%
\bibitem{malley2003}%
  \BibitemOpen
  \bibfield{author}{%
  \bibinfo {author} {\bibfnamefont{B.}~\bibnamefont{O'Malley}}\ and\ \bibinfo
  {author} {\bibfnamefont{I.}~\bibnamefont{Snook}},\ }%
  \bibfield{journal}{%
  \bibinfo {journal} {Phys. Rev. Lett.}\ }%
  \textbf{\bibinfo {volume} {90}},\ \bibinfo {pages} {085702} (\bibinfo {year}
  {2003})%
  \bibAnnoteFile{NoStop}{malley2003}%
\bibitem{karayiannis2011}%
  \BibitemOpen
  \bibfield{author}{%
  \bibinfo {author} {\bibfnamefont{N.~C.}\ \bibnamefont{Karayiannis}}, \bibinfo
  {author} {\bibfnamefont{R.}~\bibnamefont{Malshe}}, \bibinfo {author}
  {\bibfnamefont{M.}~\bibnamefont{Kr{\"{o}}ger}}, \bibinfo {author}
  {\bibfnamefont{J.}~\bibnamefont{de~Pablo}},\ and\ \bibinfo {author}
  {\bibfnamefont{M.}~\bibnamefont{Laso}},\ }%
  \bibfield{journal}{%
  \bibinfo {journal} {Soft Matter}\ }%
  \textbf{\bibinfo {volume} {8}},\ \bibinfo {pages} {844} (\bibinfo {year}
  {2012})%
  \bibAnnoteFile{NoStop}{karayiannis2011}%
\bibitem{anikeenko2007}%
  \BibitemOpen
  \bibfield{author}{%
  \bibinfo {author} {\bibfnamefont{A.}~\bibnamefont{Anikeenko}}, \bibinfo
  {author} {\bibfnamefont{N.}~\bibnamefont{Medvedev}}, \bibinfo {author}
  {\bibfnamefont{A.}~\bibnamefont{Bezrukov}},\ and\ \bibinfo {author}
  {\bibfnamefont{D.}~\bibnamefont{Stoyan}},\ }%
  \bibfield{journal}{%
  \bibinfo {journal} {J. of Non-Crystalline Solids}\ }%
  \textbf{\bibinfo {volume} {353}},\ \bibinfo {pages} {3545} (\bibinfo {year}
  {2007})%
  \bibAnnoteFile{NoStop}{anikeenko2007}%
\bibitem{bagley1970}%
  \BibitemOpen
  \bibfield{author}{%
  \bibinfo {author} {\bibfnamefont{B.}~\bibnamefont{Bagley}},\ }%
  \bibfield{journal}{%
  \bibinfo {journal} {J. Cryst. Growth}\ }%
  \textbf{\bibinfo {volume} {6}},\ \bibinfo {pages} {323} (\bibinfo {year}
  {1970})%
  \bibAnnoteFile{NoStop}{bagley1970}%
\bibitem{mcmethod}%
  \BibitemOpen
  \bibfield{author}{%
  \bibinfo {author} {\bibfnamefont{N.}~\bibnamefont{Karayiannis}}\ and\
  \bibinfo {author} {\bibfnamefont{M.}~\bibnamefont{Laso}},\ }%
  \bibfield{journal}{%
  \bibinfo {journal} {Macromolecules}\ }%
  \textbf{\bibinfo {volume} {41}},\ \bibinfo {pages} {1537} (\bibinfo {year}
  {2008})%
  \bibAnnoteFile{NoStop}{mcmethod}%
\bibitem{karayiannis2009}%
  \BibitemOpen
  \bibfield{author}{%
  \bibinfo {author} {\bibfnamefont{N.~C.}\ \bibnamefont{Karayiannis}}, \bibinfo
  {author} {\bibfnamefont{K.}~\bibnamefont{Foteinopoulou}},\ and\ \bibinfo
  {author} {\bibfnamefont{M.}~\bibnamefont{Laso}},\ }%
  \bibfield{journal}{%
  \bibinfo {journal} {Phys. Rev. Lett.}\ }%
  \textbf{\bibinfo {volume} {103}},\ \bibinfo {pages} {045703} (\bibinfo {year}
  {2009})%
  \bibAnnoteFile{NoStop}{karayiannis2009}%
\bibitem{nagel2009}%
  \BibitemOpen
  \bibfield{author}{%
  \bibinfo {author} {\bibfnamefont{L.-N.}\ \bibnamefont{Zou}}, \bibinfo
  {author} {\bibfnamefont{X.}~\bibnamefont{Cheng}}, \bibinfo {author}
  {\bibfnamefont{M.}~\bibnamefont{Rivers}}, \bibinfo {author}
  {\bibfnamefont{H.}~\bibnamefont{Jaeger}},\ and\ \bibinfo {author}
  {\bibfnamefont{S.}~\bibnamefont{Nagel}},\ }%
  \bibfield{journal}{%
  \bibinfo {journal} {Science}\ }%
  \textbf{\bibinfo {volume} {326}},\ \bibinfo {pages} {408} (\bibinfo {year}
  {2009})%
  \bibAnnoteFile{NoStop}{nagel2009}%
\bibitem{raopolymer}%
  \BibitemOpen
  \bibfield{author}{%
  \bibinfo {author} {\bibfnamefont{H.~R.}\ \bibnamefont{Vutukuri}}, \bibinfo
  {author} {\bibfnamefont{A.~F.}\ \bibnamefont{Demir{\"{o}}rs}}, \bibinfo
  {author} {\bibfnamefont{B.}~\bibnamefont{Peng}}, \bibinfo {author}
  {\bibfnamefont{P.~D.~J.}\ \bibnamefont{van Oostrum}}, \bibinfo {author}
  {\bibfnamefont{A.}~\bibnamefont{Imhof}},\ and\ \bibinfo {author}
  {\bibfnamefont{A.}~\bibnamefont{van Blaaderen}},\ \bibinfo {pages} {to be
  published}}%
   (\bibinfo {year} {2011})%
  \bibAnnoteFile{NoStop}{raopolymer}%
\bibitem{scale1}%
  \BibitemOpen
  \bibfield{author}{%
  \bibinfo {author} {\bibfnamefont{K.}~\bibnamefont{Foteinopoulou}}, \bibinfo
  {author} {\bibfnamefont{N.~C.}\ \bibnamefont{Karayiannis}}, \bibinfo {author}
  {\bibfnamefont{M.}~\bibnamefont{Laso}}, \bibinfo {author}
  {\bibfnamefont{M.}~\bibnamefont{Kr\"oger}},\ and\ \bibinfo {author}
  {\bibfnamefont{M.~L.}\ \bibnamefont{Mansfield}},\ }%
  \bibfield{journal}{%
  \bibinfo {journal} {Phys. Rev. Lett.}\ }%
  \textbf{\bibinfo {volume} {101}},\ \bibinfo {pages} {265702} (\bibinfo {year}
  {2008})%
  \bibAnnoteFile{NoStop}{scale1}%
\bibitem{scale2}%
  \BibitemOpen
  \bibfield{author}{%
  \bibinfo {author} {\bibfnamefont{M.}~\bibnamefont{Laso}}, \bibinfo {author}
  {\bibfnamefont{N.}~\bibnamefont{Karayiannis}}, \bibinfo {author}
  {\bibfnamefont{K.}~\bibnamefont{Foteinopoulou}}, \bibinfo {author}
  {\bibfnamefont{M.}~\bibnamefont{Mansfield}},\ and\ \bibinfo {author}
  {\bibfnamefont{M.}~\bibnamefont{Kr\"oger}},\ }%
  \bibfield{journal}{%
  \bibinfo {journal} {Soft Matter}\ }%
  \textbf{\bibinfo {volume} {5}},\ \bibinfo {pages} {1762} (\bibinfo {year}
  {2009})%
  \bibAnnoteFile{NoStop}{scale2}%
\bibitem{rapaportpaper}%
  \BibitemOpen
  \bibfield{author}{%
  \bibinfo {author} {\bibfnamefont{D.}~\bibnamefont{Rapaport}},\ }%
  \bibfield{journal}{%
  \bibinfo {journal} {J. Phys. A: Math. Gen.}\ }%
  \textbf{\bibinfo {volume} {11}},\ \bibinfo {pages} {L213} (\bibinfo {year}
  {1978})%
  \bibAnnoteFile{NoStop}{rapaportpaper}%
\bibitem{hartmannjcp}%
  \BibitemOpen
  \bibfield{author}{%
  \bibinfo {author} {\bibfnamefont{C.}~\bibnamefont{Hartmann}}, \bibinfo
  {author} {\bibfnamefont{C.}~\bibnamefont{Sch{\"{u}}tte}}, \bibinfo {author}
  {\bibfnamefont{G.}~\bibnamefont{Kalibaeva}}, \bibinfo {author}
  {\bibfnamefont{M.~D.}\ \bibnamefont{Pierro}},\ and\ \bibinfo {author}
  {\bibfnamefont{G.}~\bibnamefont{Ciccotti}},\ }%
  \bibfield{journal}{%
  \bibinfo {journal} {J. Chem. Phys.}\ }%
  \textbf{\bibinfo {volume} {130}},\ \bibinfo {pages} {144101} (\bibinfo {year}
  {2009})%
  \bibAnnoteFile{NoStop}{hartmannjcp}%
\bibitem{rapaportmd}%
  \BibitemOpen
  \bibfield{author}{%
  \bibinfo {author} {\bibfnamefont{D.~C.}\ \bibnamefont{Rapaport}},\ }%
  \emph{\bibinfo {title} {The Art of Molecular Dynamics Simulation}}\ (\bibinfo
  {publisher} {Cambridge University Press},\ \bibinfo {year} {2004})%
  \bibAnnoteFile{NoStop}{rapaportmd}%
\bibitem{LSA}%
  \BibitemOpen
  \bibfield{author}{%
  \bibinfo {author} {\bibfnamefont{B.~D.}\ \bibnamefont{Lubachevsky}}\ and\
  \bibinfo {author} {\bibfnamefont{F.~H.}\ \bibnamefont{Stillinger}},\ }%
  \bibfield{journal}{%
  \bibinfo {journal} {J. Stat. Phys.}\ }%
  \textbf{\bibinfo {volume} {60}},\ \bibinfo {pages} {561} (\bibinfo {year}
  {1990})%
  \bibAnnoteFile{NoStop}{LSA}%
\bibitem{karayiannis2010}%
  \BibitemOpen
  \bibfield{author}{%
  \bibinfo {author} {\bibfnamefont{N.~C.}\ \bibnamefont{Karayiannis}}, \bibinfo
  {author} {\bibfnamefont{K.}~\bibnamefont{Foteinopoulou}}, \bibinfo {author}
  {\bibfnamefont{C.~F.}\ \bibnamefont{Abrams}},\ and\ \bibinfo {author}
  {\bibfnamefont{M.}~\bibnamefont{Laso}},\ }%
  \bibfield{journal}{%
  \bibinfo {journal} {Soft Matter}\ }%
  \textbf{\bibinfo {volume} {6}},\ \bibinfo {pages} {2160} (\bibinfo {year}
  {2010})%
  \bibAnnoteFile{NoStop}{karayiannis2010}%
\bibitem{bookdoi}%
  \BibitemOpen
  \bibfield{author}{%
  \bibinfo {author} {\bibfnamefont{M.}~\bibnamefont{Doi}}\ and\ \bibinfo
  {author} {\bibfnamefont{S.}~\bibnamefont{Edwards}},\ }%
  \emph{\bibinfo {title} {The Theory of Polymer Dynamics}}\ (\bibinfo
  {publisher} {Oxford University Press},\ \bibinfo {year} {1986})%
  \bibAnnoteFile{NoStop}{bookdoi}%
\bibitem{supinf}%
  \BibitemOpen
  \bibinfo {note} {See Supplemental Material for the plot of long time
  diffusion coefficients of hard-sphere polymers.}%
  \bibAnnoteFile{Stop}{supinf}%
\bibitem{ran2010jcp}%
  \BibitemOpen
  \bibfield{author}{%
  \bibinfo {author} {\bibfnamefont{R.}~\bibnamefont{Ni}}\ and\ \bibinfo
  {author} {\bibfnamefont{M.}~\bibnamefont{Dijkstra}},\ }%
  \bibfield{journal}{%
  \bibinfo {journal} {J. Chem. Phys.}\ }%
  \textbf{\bibinfo {volume} {134}},\ \bibinfo {pages} {034501} (\bibinfo {year}
  {2011})%
  \bibAnnoteFile{NoStop}{ran2010jcp}%
\bibitem{bop}%
  \BibitemOpen
  \bibfield{author}{%
  \bibinfo {author} {\bibfnamefont{P.~J.}\ \bibnamefont{Steinhardt}}, \bibinfo
  {author} {\bibfnamefont{D.~R.}\ \bibnamefont{Nelson}},\ and\ \bibinfo
  {author} {\bibfnamefont{M.}~\bibnamefont{Ronchetti}},\ }%
  \bibfield{journal}{%
  \bibinfo {journal} {Phys. Rev. B}\ }%
  \textbf{\bibinfo {volume} {28}},\ \bibinfo {pages} {784} (\bibinfo {year}
  {1983})%
  \bibAnnoteFile{NoStop}{bop}%
\bibitem{lechner}%
  \BibitemOpen
  \bibfield{author}{%
  \bibinfo {author} {\bibfnamefont{W.}~\bibnamefont{Lechner}}\ and\ \bibinfo
  {author} {\bibfnamefont{C.}~\bibnamefont{Dellago}},\ }%
  \bibfield{journal}{%
  \bibinfo {journal} {J. Chem. Phys.}\ }%
  \textbf{\bibinfo {volume} {128}},\ \bibinfo {pages} {114707} (\bibinfo {year}
  {2008})%
  \bibAnnoteFile{NoStop}{lechner}%
\bibitem{malanoski}%
  \BibitemOpen
  \bibfield{author}{%
  \bibinfo {author} {\bibfnamefont{A.}~\bibnamefont{Malanoski}}\ and\ \bibinfo
  {author} {\bibfnamefont{P.}~\bibnamefont{Monson}},\ }%
  \bibfield{journal}{%
  \bibinfo {journal} {J. Chem. Phys.}\ }%
  \textbf{\bibinfo {volume} {107}},\ \bibinfo {pages} {6899} (\bibinfo {year}
  {1997})%
  \bibAnnoteFile{NoStop}{malanoski}%
\end{thebibliography}
%
\end{document}